\documentclass{aa}

\usepackage{graphicx}

\begin{document}
\thesaurus{03(11.19.3; 11.19.1; 11.19.2; 11.05.2; 11.09.2; 11.12.2)}

\title{Effects of galaxy mergers on the faint IRAS source counts and the 
background}

\author{Yiping Wang \inst{1,2,3,4} \and Peter L.\ Biermann \inst{3}  }
\institute{Purple Mountain Observatory, Academia
   Sinica, 210008 Nanjing, China \and National Astronomical
 observatories, Chinese Academy of Sciences \and
 Max-Planck-Institut f\"ur Radioastronomie, Auf dem H\"ugel
 69, D-53121 Bonn, Germany \and Bergische Universit\"at Wuppertal,
     D-42097 Wuppertal, Germany}
\offprints{ypwang@pmo.ac.cn}
\date{Received ??? , Accepted ???}
 \maketitle\markboth{Wang et al.: Galaxy evolution and the
IR background}{Wang et al.: Galaxy evolution and the IR
background}

\begin{abstract}
Binary merging constitutes a complementary mode of galaxy
evolution to the canonical hierarchical clustering
theory. This merger-driven evolution not only influences the
galaxy mass distribution function, but may drive 
the main galactic 
activity cycles, such as starbursts and the activity 
in the nuclei (AGNs). 
In this paper, we use galaxy aggregation dynamics together with a possible
merger-driven starburst and AGN phenomena, to study the
effects of evolution of these consequential activities for the
faint IRAS source counts and the infra-red background. We find that 
the strong evolution of IRAS 60$\,\mu m$ source count
at flux range of $10\,mJy \,\sim
\, 1\,Jy$ is difficult to be explained 
only by the merger rate decrease with cosmic time. We need to assume
a redshift dependent infrared burst phase of ultraluminous infrared
galaxies (ULIGs) from the gas rich mergers at high redshift
and the gas poor mergers at low redshift. The background intensity at
60$\,\mu m$ which we get from this aggregation evolution is
a lower limit of only $1.9\,nW\,m^{-2}\,sr^{-1}$, about half of those
estimated by some previous models, but close to the lower end of the range
derived by Malkan \& Stecker (1998).

\end{abstract}

\section{Introduction}

Over the last few years, observational progress in
discovering the faint Universe at ultraviolet (UV)/optical
and infrared wavelength via the groundbased redshift
surveys, Hubble Space Telescope (HST) multi-color images and
the infrared satellites such as IRAS (Infrared Astronomical
Satellite), ISO (Infrared Space Observatory) and COBE
(Cosmic Background Explorer), challenges and renews our
present understanding of cosmological structure formation
and the evolution of luminous galaxies, which are usually
based on two competing scenarios ``traditional monolithic
collapsing'' versus ``hierarchical clustering theories''.
The main difference between these two kinds of models is the
formation mechanism and the first epoch of the formation of 
massive galaxies.

More recently it has been shown that the emission history
(e.g. Madau plot) derived from UV/optical detections agrees quite well with the
star formation history indicated by hierarchical clustering
theories (Madau 1997, Madau et al. 1996, 1998, Ellis et al. 1996). Since the
FIR/submm unveiled a dust-shrouded early active phase and
the recent detection of high redshift evolving galaxies by
deep NICMOS/VLT images (Lilly et al. 1998, Benitez et
al. 1998), the discussion of the structure formation and the
scheme of galaxy evolution is going to require more subtlety and sophistication. For
a clear picture of the star formation history and the galaxy
evolution, we probably need more information at high
redshift, especially from FIR/submm deep surveys.

A successful interpretation of the excess of faint blue
galaxies within the merger-trigger starburst scenario in a
hierarchical clustering universe has been presented by Cavaliere
\& Menci (1997) using binary aggregation dynamics.

The binary aggregation dynamical approach for the galaxy
evolution presented by Cavaliere \& Menci (1993, 1997),
which includes more dynamics to describe a further step in
galaxy-galaxy interactions within the scheme of direct
hierarchical clustering (DHCs), probably can help to
alleviate some intrinsic problems in the DHC scenario, such
as the overproduction of small objects and the difficulty of a
reconciliation between the excess faint counts and the flat
local luminosity function.  While, on the other hand, binary
merging which plays a different and complementary role in
structure formation, can actually continue to flatten the
mass distribution function $N(m,z)$ and hence the luminosity
function $N(L,z)$ till even moderate redshift $z<1$.

Although there are many other possible evolutionary scenarios which
could interpret the present observations in this or that way,
the reason that we are encouraged to explore here a
merger-driven galaxy evolution picture with the binary
aggregation dynamics, is simply because the IRAS database
does show that most of the luminous infrared sources are
actually interacting/merging systems (Kleinmann
\& Keel 1987, Sanders \& Mirabel 1996, Hutchings \& Neff 1987, Vader
\& Simon 1987a,b), and many merger events can even occur till
moderate redshift ($z<1$).

In this paper, starting from the basic picture that galaxy
evolution is driven by the galaxy-galaxy interaction
complementary to the standard hierarchical clustering
scenario (DHCs), we investigate a simple model with the
basic concepts: 1) galaxy-galaxy interaction can even occur
till quite moderate redshifts ($z<1$) within large scale
structures that could offer the best combination of volume
and density contrast, and it could erase the dwarf part of
the mass distribution and produce the massive tail, thus
flatten the luminosity functions; 2) starburst/AGN
activities may be triggered by the merger events during
the structure formation and evolution. This is actually the
point different from other models where only the starburst
is considered as the consequence of merger events (Cavaliere
\& Menci 1997, Somerville \& Primack 1998, Somerville et al. 1998); 3) the gas rich
merger events at high redshift may trigger drastic
starburst and AGN activities in the central region of the
merging galaxies and thus enhance dramatically the FIR
luminosity because of the accumulating of dust grains from
their progenitor galaxies as well as the dust newly formed
in the star forming regions. This FIR luminosity burst phase
is usually called ultraluminous infrared galaxies (ULIGs);
4) in our model, we assume a redshift dependent infrared burst phase
for the gas rich merger events at high redshift and the gas
poor mergers at low redshift, which means the enhancement of
the infrared luminosities by the high redshift gas rich
mergers is much higher than those of low redshift gas poor
mergers. The motivation for this kind of thinking is: 1)
direct observation of enhanced starburst-AGN activities in
interacting galaxies, especially the ULIGs. This extremely
infrared bright burst phase is believed due to the starburst
merger events with far infrared luminosity $L_{fir}$
enhanced both by the accumulation of the dust mass $M_{d}$
and the increase of dust temperation $T_{d}$ as the relation
$L_{fir} \propto M_{d}\,\,T_{d}^5$. This burst phase
could increase the infrared luminosity by a factor of about 20
over that of normal starburst galaxies. (Ashby et al. 1992,
Terlevich \& Melnick 1985, Heckman 1997, Perry \&
Williams 1993, Taniguchi \& Ohyama 1998); 2) numerical 
simulation of the starburst/AGN evolutions by galaxy mergers, 
especially for their correlation and the burst phase 
(Wang \& Biermann 1998, 1999, Wang 1999); 3) the successful 
interpretation of the excess faint blue counts by galaxy 
merging and the consequential starburst activities 
(Rocca-Volmerange \& Guiderdoni 1990, Cavaliere \& Menci 
1997, Carlberg 1992, Calberg \& Charlot 1992); 4) the pair production absorption 
of high energy $\gamma$ rays by intergalactic low energy 
photons is expected to produce a high energy cutoff in 
extragalactic $\gamma$ ray spectra. The new 
data of Mkn501 appear to show an extension of the TeV 
$\gamma$ ray spectrum till about 24 TeV, which sets an upper 
limit for the intergalactic infrared emission history during 
the structure formation (Aharonian et al. 1999).

Since the dust-shrouded geometry could strongly affect the
AGN spectrum and cause significant radiation at infrared
wavelength in the framework of a Unification Scheme, we thus
add the statistics of nuclear activity in galaxies as one additional
constraint in the context of the various models already in the literature.
We know that nuclear activity is detected to very high redshift 
and provides a strong constraint on the cosmological evolution models.  
Therefore we are less
conservative than Malkan \& Stecker (1998), but use more constraints than,
e.g., Somerville \& Primack (1998).  Obviously, any such modelling depends on
the basic concepts used, and thus our results provide a strong lower limit to
the far-infrared background.

The outline of this paper is as follows: 1) in Sect.
\ref{dyn}, we introduce the binary aggregation dynamics, and
the models of the interaction kernel; 2) we discuss the prescriptions
of mass-to-light ratio in our model for the luminous
infrared galaxies which is enhanced by the starburst and AGN
activities in Sect. \ref{mass-light}, especially modelling of the
redshift dependent infrared burst phase for ultraluminous infrared galaxies (ULIGs)
from gas rich mergers at high redshift and a suppressed
burst phase for gas poor mergers at low redshift; 3) in
Sect. \ref{num}, we discuss the
numerical simulations and compare the Monte Carlo results
with the IRAS $60\,\mu m$ source counts from three major
infrared sources (starburst galaxies, Seyferts and
spirals). We check also the redshift distribution of these
infrared sources which are brighter than $S_{60}\sim
10\,mJy$, calculate the integrated background level at
wavelength of $60\,\mu m$, and make an extrapolation to $25\,\mu
m$ and $100\,\mu m$ based on the model spectrum and souce counts. Finally, we give our conclusions.

\section{Binary aggregation theory\label{dyn}}

The classical approach to aggregation phenomena is based on
the Smoluchowski(1916) equation. We start with the continuous form:

\begin{eqnarray}
\label{von}
&&\frac{\partial\rm N(m,t)}{\partial t}=\nonumber\\
&&\frac{1}{2}\int_{m_{\ast}}^{m-m_{\ast}}dm^{'}K(m^{'},m-m^{'},t)
N(m^{'},t)N(m-m^{'},t)\nonumber\\
&&-N(m,t)\int_{m_{\ast}}^{m_{\ast\ast}-m_{\ast}}dm^{'}K(m,m^{'},t)N(m^{'},t)
\end{eqnarray}

\noindent where $N(m,t)$ is the mass distribution function in
``comoving'' form, which describes the number density of
galaxies within the mass range ($m$, $m+dm$) at cosmic time
$t$. The mass $m \equiv M/M_{\ast}$ is
normalized in terms of $M_{\ast}$, which is the mass
corresponding locally to the standard characteristic
luminosity $L_{\ast}$ (see Peebles 1993). $m_{\ast}$ and
$m_{\ast\ast}$ represent the lower and upper limits on the
masses of galaxies. Usually we can set $m_{\ast} = 0,
m_{\ast\ast} = \infty$ in case when $m_{\ast}$ is actually
much smaller than the total mass of considered system.

The kernel $K(m,m^{'},t)=n_{g}(t)<\Sigma V(t)>$ reflects the 
interaction rate for each pair of masses $(m, m^{'})$, and 
depends on the mechanism and environment of such 
aggregations.  $n_{g}$ is the density of galaxies, it 
scales on average with the expansion of Universe as $n_{g} 
\propto (1+z)^3$; $V(t)$ is the relative velocity of the 
interacting pairs, and $\Sigma$ is the cross section. The 
average of them runs over the galaxy velocity 
distribution.

In aggregation dynamics, the interaction kernel $K =
n_{g}<\Sigma V>$ is the key point for driving the whole
evolutionary process, which depends strongly on the
environment structures.

In this context, we discuss simply the case when the
colliding galaxies are in certain structures, thus the cross
section could be assumed to be the hyperbolic encounter
prescription by Saslaw (1985), Cavaliere et al. (1991, 1992), which is in the form:

\begin{equation}
\Sigma =
\epsilon\left(\frac{V}{v_{m}}\right)\pi\left(r_{m}^2+r_{m^{'}}^2\right)\left
[ 1+ \frac{G(m+m^{'})}{r_{m}V^2}\right]
\end{equation}

\noindent with $r_{m}=r_{\ast}m^{1/3}, v_{m}=v_{\ast}m^{1/3}$, where
$v_{\ast}$ and $r_{\ast}$ are the velocity dispersion and
the dark halo radius of the present $M_{\ast}$ galaxy. The
function $\epsilon\left(\frac{V}{v_{m}}\right)$ describes
the decreasing efficiency of the aggregations with an
increasing relative velocities, which could be determined by
the N-body simulations (see, Richstone \& Malumuth 1983).

Because of the uncertainty and the complexity of the
components in the interaction kernel, such as the average
density of galaxies in the environment $n_{g}(t)$, the
relative velocity distribution of the aggregation pairs
$V(t)$ and the interaction cross section $\Sigma$, the exact
prescription of the interaction kernel is still poorly
known. We will thus adopt in our simulation a simplified
formula for the aggregation kernel $K(m,m^{'},t)$ with 
separated time evolution term and the cross section
term. We give the interaction kernel $K(m,m^{'},t)$ as:

\begin{equation}
\label{kmk}
K(m,m^{'},t) \propto
t^{k}(m^{2/3}+{m^{'}}^{2/3})\left[1+\alpha(m^{2/3}+{m^{'}}^{2/3})\right]
\end{equation}

\noindent where $t^{k}$ is the cosmic time evolution term. The free
parameter $k$ depends on the specific structures like
clusters, filaments or sheets; $\alpha$ is a free parameter in
our model, which describes the relative importance of 
two kinds of encountering (geometric collisions and focused interactions) in the cross section.
The aggregation time scale is $\tau$, and thus the
aggregation rate is $\tau^{-1}$, which is proportional to
the interaction kernel $K(m,m^{'},t)$ as $\tau^{-1}\propto
K(m,m^{'},t)$.

\section[Modelling the ULIGs at high redshift]{Modelling the
ultraluminous infrared phase at high
redshift\label{mass-light}}

The binary aggregation dynamics and Monte Carlo
simulations give us information about the evolution of the
galaxy mass distribution function $N(m,t)$. What we
can actually observe and use to constrain the galaxy
evolution model is: 1) the luminosity functions of galaxies
in general or with certain morphologies; 2) the source
counts, redshift distributions and the background intensity
from various Hubble Deep Field surveys and the
background measurements.

In this section, we will discuss the conversion
of the mass distribution function $N(m,t)$ by Monte
Carlo simulation of a merger-driven galaxy evolution
scenario to the observable luminosity function $N(L,t)$.

In this case, we
need to know i) a simple prescription of mass-to-light ratio
for starburst galaxies and AGNs; and ii) the bolometric
correction for certain spectral characteristics, especially at
infrared wavelengths.

A simple discussion of the mass-to-light ratio for the
starburst galaxies is given by Cavaliere \& Menci (1997) for
the faint blue galaxies who estimate the luminosity of
starburst galaxies by the gaseous mass of the galaxies and
the dynamical time scales.  The mass to light relation of
those blue starburst galaxies is given as:

\begin{equation}
\label{ml}
\frac{L}{L_{\ast}} = \left ( \frac{M}{M_{\ast}}\right )^\eta
\hspace{0.5cm} and \hspace{0.5cm}L_{\ast}(z)\propto
f(z,\lambda_{0},\Omega_{0})
\end{equation}

\noindent where $\eta = 4/3$; $L$ is the bolometric luminosity and $M$ is
the mass of the galaxy; $L_{\ast}$ is the local standard
characteristic luminosity, with the corresponding mass
$M_{\ast}$ (Peebles 1993).  $L_{\ast}(z)\propto
f(z,\lambda_{0},\Omega_{0})$ represents a cosmological
redshift dimming. Considering the large scale structure,
Cavaliere et al. (1997) gave a prescription of $f(z)$ being
a function of dimensionality $D$ of the large scale
structure as $L_{\ast}\propto f(z)\propto (1+z)^{(3+D)/2}$.
Studies of the origin of ultraluminous infrared galaxies
such as Arp220, NGC1614, NGC3256 and IRAS18293-3413, 
show that the infrared
luminosity of this kind of galaxies is about
a factor of 20 higher than that of
normal starburst galaxies, with a statistical relation of
$L_{fir}\propto M_{d}\,T_{d}^5$ (Taniguchi \& Ohyama
1998). The increase in both the dust mass $M_{d}$ and the
dust temperature $T_{d}$ by starburst merger events could
enhance the far infrared luminosities of starburst merging
galaxies dramatically; we thus call them ultraluminous
infrared galaxies (ULIGs).  Because of the uncertainties of
the gas/dust ratio and the complexity of the consequential
heating of the dust grains, the mass-infrared luminosity
correlation for the ultraluminous infrared galaxies is
still unclear. In our calculation, we thus simply adopt the
power law relation for the mass-to-infrared luminosity
similar to that of starburst galaxies by Cavaliere \& Menci (1997), with the exponent $\eta$ increased by a factor
of about two to simulate the situation of an enhancement of
their infrared luminosity by about a factor of near 20 for a
typical ULIGs of mass $\sim 10^{12}\,M_{\odot}$.

In our model, we basically assume that starburst and AGN
activities which are triggered by the merger events at high
redshift are more drastic than those of low redshift, simply
because the mergers are usually between gas rich systems at
high redshift and the progenitors of the low redshift
mergers are already poor in cold gas. In this case, we
assume in our model a redshift depenent infrared burst phase, which
means the enhancement of infrared radiation from gas rich
mergers at high redshift is much higher than that of low
redshift.  We simulate this effect with a power law
suppression of the infrared burst luminosity simply as
$L_{ir}(z-\triangle z)\propto L_{ir}(z)^{-\zeta}$ below a
transition redshift $z \sim 1$, besides a normal redshift
dimming defined in Eq. (\ref{ml}). This power law
suppression means that the infrared luminous galaxies at the
bright tail of the luminosity function become gas poor
faster than the less luminous ones. On the other hand, choosing redshift $z\sim 1$ as a
redshift transition for the gas rich mergers at high
redshift and gas poor mergers at low redshift is based on
the consideration that the cosmic time scale of $z\sim 1$ is
about $3 \times 10^9$ years, which is 
approximately the time scale for the disk
evolution (Lin \& Pringle 1987), it probably indicates a
stage when galaxies are becoming gas poor.

While the basic concept is clearly correct, the details of
these assumptions are a bit crude and arbitrary, especially the assumptions
for the transition redshift $z\sim 1$ and the consequential
power law suppression of the infrared burst phase. But it seems that they are
the important effects for the fitting of IRAS source
counts in our simulation, especially the strong evolution
at flux range $10\,mJy \,\sim \, 1\,Jy$. Varying any of the model
parameters could influence more or less the evolution of the luminosity functions and the source
counts finally. While all of the model parameters combine to influence
the luminosity functions together, the impact from variation of one of the parameters such as $\alpha,\,\,D$ or $\eta$, probably still could be compensated by varying again 
other related
parameters. As for the assumptions of transition redshift and the differential burst phase, there seems to be not much room
left for the variation of these particular values. We checked the case when the transition redshift and the enhancement
of the infrared burst phase $\eta$ are both twice and half of the values in our model. Although we adjusted other
parameters in order to approximately fit the observed local luminosity functions for the infrared luminous sources,
it seems that they are far from giving an acceptable fit to the IRAS deep surveys (see Fig.\ref{db}). In this case, what we show here is only 
one of the most possible, realistic set of model parameters which could
give the best fitting of available observations. Since our model is still too crude, we need probably further information about
the evolution of infrared bright sources at high redshift,
or the redshift distributions to high redshift, to give more robust
model constraints. We will discuss this point again later.

\section{Numerical simulation and discussion\label{num}}

We use a Monte Carlo inverse-cascading process to simulate
the binary aggregation evolution of galaxies which is
described by Eq. (\ref{von}).  In our simulation, we
consider the initial galaxy mass distribution as a
$\delta$-function starting from redshift $z=15$ and of mass $M=2.5 \cdot 10^{10} M_{\odot}$ as an approximation. Since
binary aggregation dynamics does not strongly depend on
the initial condition, the memory of the initial
condition will disappear after the transients and the mass
distribution is independent of initial details with
self-similar evolution.  The specific discussion of this process was
presented by Cavaliere et al. (1991, 1992) analytically and
numerically.

\begin{figure}
\label{fcount}
\resizebox{\hsize}{!}{\includegraphics[]{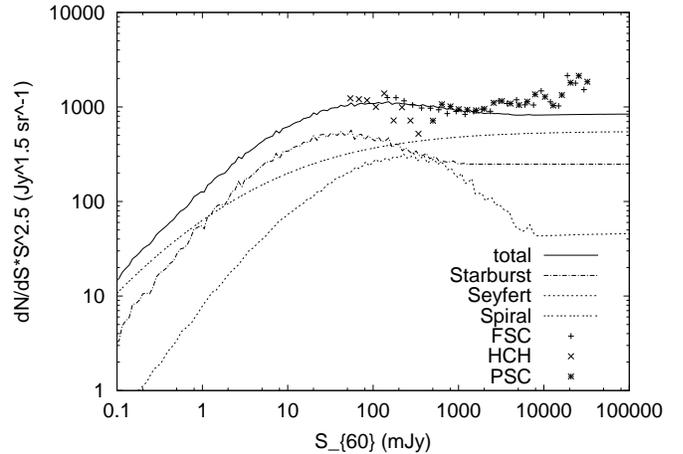}}

\caption{\label{fcount} This is our fit to the IRAS $60\,\mu m$
source counts using three major populations (starburst
galaxies, Seyferts and spirals).  The data are from IRAS
Point Source Catalog (1985) with galactic latitudes $|b|\geq
50^{\circ}$ (PSC), Hacking et al. IRAS deep survey (HCH),
FSC from deep surveys by Moshir et al. (1992) and Saunders
(1990). We include here also the data by Gregorich et
al. (1995) (GNS). It seems that the GNS data could
overpredict the IR background and probably their sample has
a biased detection (Bertin et al. 1997).  The source counts
of starburst galaxies and Seyferts are from the Monte Carlo
simulation where the evolution of both activities are
triggered by the galaxy-galaxy interactions/mergers during
the structure formation and evolution. The spirals are
assumed to have an almost constant star formation history
since their formation. The wiggles in the curve are from the
numerical inaccuracy of the binning during the Monte Carlo
simulation. We consider the excess of the data of $\sim
10\,Jy$ to be a cosmologically local phenomenon, since we
are inside a local sheet.}
\end{figure}

In this case, evolution of the dwarf galaxies with mass less than
$\sim 10^{10} M_{\odot}$ is not relevant to our results.
Even for those dwarf galaxies of mass $\sim 10^{10}\,M_{\odot}$, we could see that
their influence is not very important since our results only strongly depend
on the evolution of infrared-luminous galaxies.

The large scale structures (LSSs), which have the advantage
of a larger density and lower velocity dispersions compared
with the field and the virialized clusters, present an
ideal environment for galaxy-galaxy interactions to
take place in. However, they are still quantitatively less
understood both from the observational and theoretical point
of view. In the paper about the evolution of faint blue galaxies, Cavaliere \& Menci (1997) simply discussed the relation of the
evolution of merger rate which is represented by the
term $t^{k}$ in Eq. (\ref{kmk}) with
different environments, such as the homogeneous ``field'',
the virialized clusters or groups and the large scale
structures (sheets and filaments). They include a free parameter $D$ in the expansion of Universe
for the Large Scale Structures.

\begin{equation}
f(z)=\left[\frac{1+z}{1+z_{in}}\right]^{(3+D)/2}
\end{equation}

\noindent where $z_{in}$ is the redshift when galaxy aggregation
becomes effective; $D$ is the dimensionality of the large
scale structure, with $D=2$ for sheetlike structures and
$D=1$ for filaments.

We can transform $f(z)$ to be a function of cosmic time $t$
with the conversion of $z$ to $t$ for a flat universe (see
Peebles 1993) as:

\begin{equation}
1+z=\left(\frac{2}{3}\,\,\frac{t^{-1}}{H_{0}\,\Omega^{1/2}}\right)^{2/3}
\end{equation}

\noindent so, we get $f(t) \propto t^{k}$, with $k$ at range of (-4/3,
-5/3) corresponding to $D=1$ (filaments) and $D=2$ (sheets).

\begin{figure}
\label{dis}
\resizebox{\hsize}{!}{\includegraphics[]{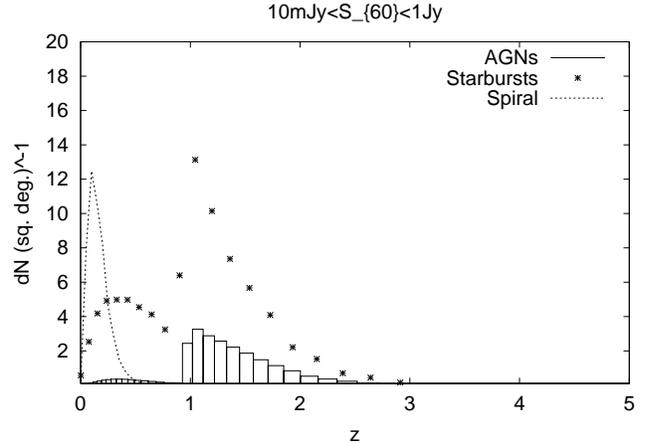}}

\caption{\label{dis} This is the redshift distribution of starburst
galaxies, Seyferts and spirals at flux density range of
$10\,mJy \,\sim
\, 1\,Jy$. The high
redshift contribution ($z \sim 1.5$) is the ultraluminous
infrared burst phase from the gas rich mergers of high
redshift. The newest result from the ISO far-infrared survey at
$175\,\mu m$ suggests that half of the FIRBACK sources are
probably at redshifts greater than 1 (Dole et al. 1998),
which gives further ``motivation'' for considering the model.}  
\end{figure}

In our simulation, the best fit structure for the luminous
infrared galaxies are more like in the filaments with $D
\sim 1$ and $k \sim -4/3$. The free parameter $\alpha$ in
equation (\ref{kmk}), which describes the relative weight of
two kinds of interactions (geometric collisions or
gravitationally focused interactions), is close to 0.9. 
But we know from our simulation, although varying these parameters could
influence the final results, we probably still could compensate such effects
with the adjustment of other relevant model parameters. So, what we show here is
only one set of the most possible values which could give the best fit to the present observations.

In our model, we assumed a redshift dependent
infrared burst phase for the merger triggered luminous infrared
sources since a transition
redshift $z \sim 1$, which means the enhancement of the far
infrared luminosity of ULIGs from merging galaxies at high
redshift is much higher than that of low redshift, simply
because the high redshift mergers are usually between gas
rich systems but between gas poor systems at low redshift.
We found in our simulation although the assumption of the
``differential burst'' phase and the consequential power law
supression is very arbitrary and crude, it is actually very
important for the fitting of the strong evolution of 
IRAS $60\,\mu m$ source counts at flux range of
$10\,mJy \,\sim \, 1\,Jy$. We show the results in Fig.\ref{fcount}, where the
starburst galaxies and Seyferts are both assumed to be of
the evolution which is driven by galaxy-galaxy
mergers/interactions, while spirals basically keep a
constant star formation history since their formation.  We
normalized our Monte Carlo simulation with the observed
local luminosity function at $60\,\mu m$ of starburst
galaxies from Saunders (1990) and Seyferts from Rush
et al. (1993). Varying this transition redshift by a factor of two already
has significant effects for the fitting of the source counts (see Fig. \ref{db}).
But, since our model is still too crude, we probably only can say what we present here
is a most probable scenario which gives best fit to the IRAS deep surveys and other available observations.

\begin{figure}
\label{db}
\resizebox{\hsize}{!}{\includegraphics[]{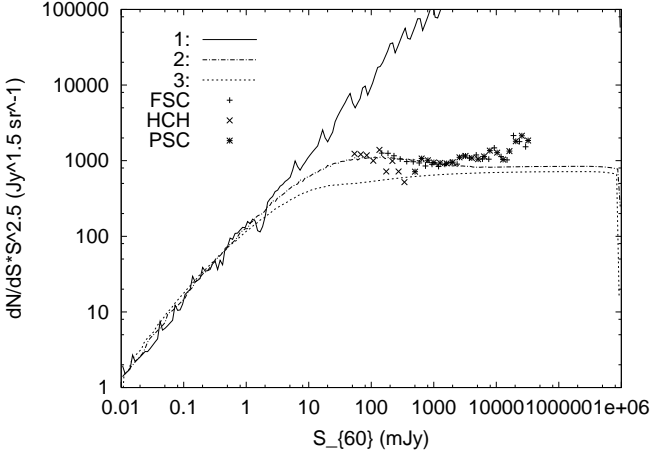}}

\caption{\label{db} This is the fitting of IRAS $60\,\mu m$
source counts for different transition redshift and the infrared burst
phase $\eta$. case 1: $z=2,\,\,\eta=4.6$; case 2: $z=1,\,\,\eta=2.3$; case 3:
$z=0.5,\,\,\eta=1.4$. We could see that case 2 is the best fit model parameters,
meanwhile this set of parameters has the best fit also to the local luminosity functions of
infrared luminous sources.}
\end{figure}

We consider in our simulation both types of AGNs with the
assumption that the abundances of Seyfert I and Seyfert II
at low redshift are approximately equal, which is suggested
by the extended $12\,\mu m$ galaxy sample (Rush et
al. 1993), and the recent Hubble Space Telescope imaging
survey of nearby AGNs (Malkan et al. 1998). At high redshift ($z>1$), we basically assume that the dust shrouded phase
dominates and has a fraction of about $80\%$ which is also
suggested by the recent cosmic X-ray background (Gilli et
al. 1999).

We choose NGC1068 IR spectra as the standard template for
all the obscured AGNs at low redshift, while all ``Type I'' have a
SED well represented by the mean SED(spectral energy distribution)
of Seyfert I. We also assume that the early phase of these AGNs show
the typical spectra such as the dust shrouded F10214+4724, and a phase
poor in cold gas like the Cloverleaf quasar. The template of all these spectra
were well modelled by Rowan-Robinson (1992), Rowan-Robinson et al. (1993) and Granto
et al. (1994, 1996, 1997).

In order to understand the 
contributing sources for the faint slope of the IRAS
$60\,\mu m$ source counts,
we plot out also the redshift distributions at flux
range of $10\,mJy \sim 1\,Jy$ for the three important
populations (starburst galaxies, Seyferts and spirals). 
We see from Fig.\ref{dis} that
this faint slope comes from the low redshift starburst
galaxies which peak at $z\sim 0.5$, the local spirals with
mean redshift about 0.1 and the infrared burst phase of high
redshift gas rich mergers which peaks at $z\sim 1.5$. Fig.\ref{dis}
can be used to make powerful predictions about the redshifts of faint
infrared sources. It seems to imply that about two-thirds of the
faint $60\,\mu m$ sources should have redshifts from a little less than
1 to 2 (and that a fair fraction of those will be Seyferts). Recent ISO and NICMOS results
of the high redshift ultraluminous infrared galaxies further increases our
motivaton to consider the model (Dole et al.
1998, Benitez et al. 1998).

\begin{figure}
\label{bkg}
\includegraphics[width=9cm, height=12cm, angle=0]{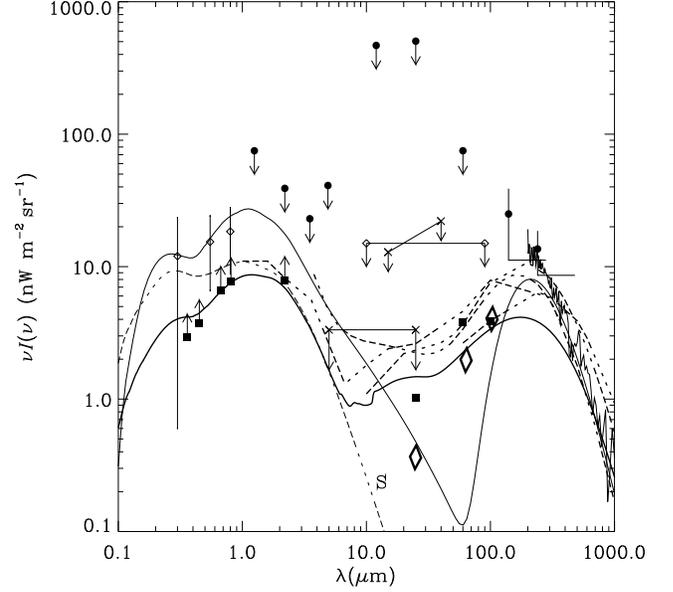}
\caption{\label{bkg} The extragalactic IR background spectrum predicted
by various evolution models and from the COBE results (Dwek et al. 1998). The
DIRBE $2\sigma$ upper limits at 1.25 - $60 \,\mu m$, and
detections at 140 and $240\,\mu m$ with $\pm 2\sigma$ error
bars are represented by solid circles (Hauser et
al. 1998). The FIRAS 125 - $5000\,\mu m$ detection (Fixsen
et al. 1997) is shown by a light dashed line; the UV-optical
lower limits of Pozzetti et al. (1997) and the $2.2\,\mu m$
lower limit (Gardner et al. 1997) are represented by solid
squares. The ``X'' represent the upper limits on the EBL
derived from TeV observations of Mrk 501 (Stanev \&
Franceschini 1998), and Mrk 421 (Dwek \& Slavin 1994), and
the open small diamonds represent the upper limits derived
from the analysis of the fluctuations the DIRBE maps
(Kashlinsky et al. 1996). The heavy solid curve represents the EBL spectrum
calculated from the model with the UV/optical indicated SFR
and the dust extinction effects (Dwek et al. 1998). Other
models presented in the figure are the Backwards Evolution
models of Malkan
\& Stecker (1998; dotted line), and of Beichman \& Helou (1991;
dashed dotted line); the Forward Evolution models of
Franceschini et al. (1997, dashed line), and Guiderdoni et
al. (1997, dashed triple dotted line); and the Cosmic
Chemical Evolution model of Fall et al. (1996, thin solid line).  The dashed line marked $S$ represents
the EBL spectrum calculated for a dust$-$free universe with
the UVO cosmic star formation rate. The lower limits from
IRAS counts at 25, 60 and $100\,\mu m$ by Hacking \& Soifer
(1991) are also represented here by the solid squares. We
plot out the revised lower limits at these wavelengths from
our model with three bigger and thicker open diamonds
``$\diamondsuit$''. }
\end{figure}

We calculate the background level $\nu I_{\nu}$ at $60\,\mu
m$, it is approximately $1.9\,nW\,m^{-2}\,sr^{-1}$.  We then
extrapolate our simulation to wavelengths of $25\,\mu m$ and
$100\,\mu m$. The background intensity are
all shown in Fig.\ref{bkg}.  Clearly, our extrapolation to 
$25\,\mu m$ is too low, compared to other models, which 
suggests that we have too low a mixture of intermediate dust 
temperatures in our templates for the emission spectra.  
However, our $60\,\mu m$ background is very close to the 
lower end of the range empirically derived by Malkan \& 
Stecker (1998), and so can be considered as a firm lower limit.

\section{Conclusion}  

We described in this paper a Monte Carlo simulation of the
inverse-cascading process of a
merging-driven galaxy evolution scenario, where the evolution
of infrared luminous starburst galaxies and AGNs may be triggered
by the galaxy-galaxy interactions till even moderate
redshift (say, $z < 1$) in the Large Scale Structures. We assume
in our model a redshift dependent infrared burst phase which is based
on the concepts that starburst and AGN activities triggered
by gas rich mergers at high redshift are more drastic than
those of low redshift, thus the enhancement of the far
infrared luminosities in these ULIGs from the high redshift
merger events are higher than that of low redshift. We
simulate this effect in our calculation by a power law
suppression of the infrared burst phase since a transition
redshift $z\sim 1$. We adopt the transition redshift here at
$z\sim 1$ simply because the cosmic time scale of $z\sim
1$ ($3\times 10^9$ years) is approximately the disk
evolution time scale (Lin \& Pringle 1987). Varying any of the model parameters could influence the evolution of
luminosity functions of infrared luminous sources and thus influence the
souce counts more or less.  But, no matter how
we adjust the relevant parameters, such as $\alpha$, $D$ and $\eta$, in
order to get a strong decrease of merger rate with cosmic time around redshift $1\sim 2$ for the source count fitting,
it seems that the quick fading and suppression of the infrared burst phase at redshift $z\sim 1$ is
a critical effect in our
simulation to interpret the strong evolution of the IRAS
$60\,\mu m$ source counts at flux range $10\,mJy\sim
\,1\,Jy$. The impact from the variation of model parameters such as $\alpha$, $D$ and $\eta$, probably could be compensated
by varying again other related parameters, thus their particular values in our model are not the critical points for such
an evolutionary scenario; However, varying the transition redshift and the infrared enhancement $\eta$ by a
factor of two could strongly influence the final results. So, it appears that there is
not much room for varying any of these parameters. Since our model is still too simple and
crude, we probably need further information about the evolution of luminosity functions
at high redshift for those infrared bright sources in order to give strong model
constraints.

We checked also the redshift distribution of the three
major infrared sources (starburst galaxies, Seyferts and
spirals). We see from Fig.\ref{dis} that the mean redshift of
starburst galaxies and AGNs which are brighter than
$S_{60}\sim 10\,mJy$ is around $z\sim 0.5$ and quickly
diminish till $z\sim 1$; a new population of high redshift
ultraluminous infrared burst phase at mean redshift $z\sim
1.5$ then takes over. This seems to be consistent with the
present groundbased NIR and optical/UV HDF surveys, where they
failed to have enough detection of starburst
galaxies near $z\sim 1$ (Ashby et al. 1992, Koo \& Kron
1992). On the other hand, recent NICMOS/VLT and FIR/submm
survey surely found a certain amount of infrared bright
galaxies at high redshift, and the newest result of FIRBACK
(Far Infrared Survey) with ISO shows that more than half of
the ultraluminous infrared galaxies are at redshift $z>1$
(Lilly et al. 1998, Benitez et al. 1998, Dole et
al. 1998). This provides further motivation to consider their contribution to the strong evolution of the IRAS faint source
counts at flux range of $10\,mJy\sim 1\,Jy$ in
Fig.\ref{fcount}.

Meanwhile the new data of nearby blazars like Mkn501 by HEGRA team, appears
to show an extension of the TeV 
$\gamma$ ray spectrum till about 24 TeV, which sets an upper 
limit for the intergalactic infrared emission history during 
the structure formation (Aharonian et al. 1999).
We calculated the background level at $60\,\mu m$ from a possible merger driven starbursts and AGNs
scheme and a simple constant star formation
history for the spiral galaxies. The infrared
background level at $60\,\mu m$ is only
$1.9\,nW\,m^{-2}\,sr^{-1}$, which is about half of those
estimated by some previous papers and consistent with the
upper limit from the new data of TeV $\gamma$ ray spectrum
of Mkn501 (Stecker 1999, Funk et al. 1998).  Clearly, this 
is a strong lower limit, because any variation on our 
model could produce a higher background.  The forth-coming 
data on direct source counts at infrared wavelengths will 
allow to further constrain the evolution of both AGN and 
starbursts, as well as the absorption at gamma rays near 10 TeV
photon energy.

\begin{acknowledgements}
We should thank Profs. Martin Harwit, Hinrich  Meyer,
Drs. Norbert Magnussen, Karl Mannheim and Wolfgang Rode for their 
helpful comments and suggestions. In addition PLB would 
like to thank Profs. Ocker de Jager, Todor Stanev, 
Floyd Stecker, Joel Primack and Amri Wandel for intense discussions 
of the theme of this work.  YPW was supported by the 
PhD fellowship of Wuppertal University of Germany and 
finished the calculations at MPIfR. YPW is indebted 
to their hospitality and kindly helps. We are very grateful to the anonymous referee for her/his
helpful comments and actually suggestions to consider the model further.

\end{acknowledgements}


\begin{thebibliography}{}

\bibitem{} Aharonian F., for the HEGRA collaboration, 1999, A\&A 349, 11, astro-ph/9903386

\bibitem{} Ashby M., Houck J.R., Hacking P.B., 1992, AJ 104, 980

\bibitem{} Beichman C. A., Helou G. 1991, ApJ 370, L1

\bibitem{} Benitez N., Broadhurst T., Bouwens R. et al., 1998, ApJL accepted, 
astro-ph/9812205

\bibitem{} Bertin E., Dennefeld M., Moshir M., 1997, A\&A 323, 685

\bibitem{} Carlberg R. G., 1992, ApJ 399, L31

\bibitem{} Carlberg R. G., Charlot S., 1992, ApJ 397, 5

\bibitem{} Cavaliere A., Menci N. 1993, ApJ 407, L9

\bibitem{} Cavaliere A., Menci N. 1997, ApJ 480, 132

\bibitem{} Cavaliere A., Colafrancesco S., Scaramella R., 1991, ApJ 380, 15

\bibitem{} Cavaliere A., Colafrancesco S., Menci N. 1992, ApJ 392, 41

\bibitem{} Dole H., Lagache G., Puget J. L. et al., 1998, ``The Universe as
seen by ISO'', UNESCO, Paris, astro-ph/9902122

\bibitem{} Dwek E., Arendt R. G., Hauser M. G. et al., 1998, ApJ 508, 106

\bibitem{} Dwek E., Slavin, J. 1994, ApJ 436,696

\bibitem{} Ellis R. S., Colless M., Broadhurst T. et al., 1996, MNRAS 280, 235

\bibitem{} Fall S. M., Charlot S., Pei Y. C. 1996, ApJ 464, L43

\bibitem{} Fixsen, D. J., Weiland J. L., Brodd S. et al., 1997, ApJ 490, 482

\bibitem{} Franceschini A., Aussel H., Bressan A. et al, 1997, Review in ESA FIRST symposium(ESA SP 401), astro-ph/9707080

\bibitem{} Funk B., Magnussen N., Meyer H. et al., 1998, Astroparticle
Physics 9, 97-103

\bibitem{} Gardner J. P., Sharples R. M.,
Frenk C. S., Carrasco B. E., 1997, ApJ 480, L99

\bibitem{} Gilli R., Risaliti G., Salvati M., 1999, A\&A 347, 424, astro-ph/9904422

\bibitem{} Granato G. L., Danese L., 1994, MNRAS 268, 135

\bibitem{} Granato G. L., Danese L., Franceschini A., 1996, ApJ 460, L11

\bibitem{} Granato G. L., Danese L., Franceschini A., 1997, ApJ 486, 147

\bibitem{} Gregorich D.T., Neugebauer G., Soifer B.T. et al., 
1995, AJ 110, 259

\bibitem{} Guiderdoni B., Bouchet F.R., Puget J.L. et al., 
1997, Nat 390, 257

\bibitem{} Hacking P. B., Soifer B. T., 1991, ApJ 367, L49

\bibitem{} Hacking P. B., Condon J. J., Houck J. R., 
1987, ApJ 316, L15 (HCH)

\bibitem{} Hauser, M.G., Arendt R.G., Kelsall T. et al., 1998, ApJ 508,
25 (Paper I)

\bibitem{} Heckman T.M., 1997, in "The Ultraviolet Universe at Low and
High Redshift", ed. W.Waller(AIP)

\bibitem{} Hutchings J.B., Neff S.G., 1987, AJ 93, 14

\bibitem{} Kashlinsky A., Mather J. C., Odenwald S., 1996, ApJ 473, L9

\bibitem{} Kleinmann S. G., Keel W. C., 1987, in ``Star Formation in
Galaxies'', ed. Lonsdale-Persson C. J., p.559. Washington
DC: US Govt.  Print. Off.

\bibitem{} Koo D.C., Kron R.G., 1992, ARA\&A 30, 613

\bibitem{} Lilly S. J., Eales, S. A., Gear W. K. et al. 1998, in ``The
Next Generation Space Telescope: Science Drivers and
Technological Challenges'', 34th Li$\grave{e}$ge
Astrophysics Colloquium, astro-ph/9807261

\bibitem{} Lin D.N.C., Pringle J.E., 1987, ApJ 320, L87

\bibitem{} Malkan M. A., Stecker F. W. 1998, ApJ 496, 13

\bibitem{} Malkan M. A., Gorjian V., Tam R., 1998, ApJS 117, 25

\bibitem{} Madau P., 1997, in ``Galaxy Interactions at low and High
Redshift'', IAU, Symp. no.186, Kyoto, Japan, 17-30 August,
1997

\bibitem{} Madau P., Ferguson H. C., Dickinson M. E. et al., 1996, MNRAS
283, 1388

\bibitem{} Madau P., Pozzetti L., Dickinson M. E., 1998, ApJ 498, 106

\bibitem{} Moshir M., Kopman G., Conrow T. A. O., 1992, Explanatory Supplement to the IRAS
Faint Source Survey, Version 2 (JPL, Pasadena, CA), D-100158/92

\bibitem{} Peebles P.J.E., 1993, in ``Principles of Physical Cosmology''
by Princeton University Press

\bibitem{} Perry J.J.,, Williams R., 1993, MNRAS 260, 437

\bibitem{} Pozzetti, L., Madau, P., Ferguson, H. C., et al.,
1997, MNRAS 298, 1133

\bibitem{} Richstone D., Malumuth E.M., 1983, ApJ 268, 30

\bibitem{} Rocca-Volmerange B., Guiderdoni B., 1990, MNRAS 247, 166

\bibitem{} Rowan-Robinson M., 1992, MNRAS 258, 787

\bibitem{} Rowan-Robinson M., Benn C. R., Lawrence A. et al., 1993, MNRAS 263, 123

\bibitem{} Rush B., Malkan M. A., Spinoglio L., 1993, ApJS 89, 1

\bibitem{} Saslaw W.C., 1985, in ``Gravitational Physics of Stellar and
Galactic Systems'' (Cambridge: Cambridge Uni.Press), 231

\bibitem{} Sanders D. B., Mirabel I. F., 1996, ARA\&A 34, 749

\bibitem{} Saunders W., 1990, PhD thesis, Statistical Cosmology with IRAS
Galaxies: The Large Scale Structure and Evolution of the
Universe (Queen Mary College, University of London)

\bibitem{} Smoluchovski M., 1916, Phys. Z. 17, 557

\bibitem{} Somerville R.S., Primack J.R., 1998, MNRAS 
in press, astro-ph/9802268

\bibitem{} Somerville R.S., Primack J.R., Faber S.M., 1998 MNRAS in press
astro-ph/9806228

\bibitem{} Stanev T., Franceschini A. 1998, ApJ 494, L159

\bibitem{} Stecker F. W., 1999, in Proc. DPF99 Conf., astro-ph/9904416

\bibitem{} Taniguchi Y., Ohyama Y., 1998, ApJ 508, L13

\bibitem{} Terlevich R., Melnick J., 1985, MNRAS 213, 841

\bibitem{} Vader J. P., Simon M. 1987a, Nat 327, 304

\bibitem{} Vader J. P., Simon M. 1987b, AJ 94, 854

\bibitem{} Wang Y.P., 1999, PhD thesis, ``Merger-driven starbursts and AGN,
cosmology and the infrared background''( Wuppertal
University, Germany)

\bibitem{} Wang Y.P., Biermann P.L., 1998, A\&A 334, 87

\bibitem{} Wang Y.P., Biermann P.L., Sept. 20-25, 1999, in the conference proceeding of 
``The $5^{\rm th}$ CAS-MPG workshop on high energy astrophysics'', 
Urumqi, China


\end{thebibliography}
\end{document}